\documentclass[copyright,creativecommons]{eptcs}
\providecommand{\event}{ICLP 2026} % Name of the event you are submitting to

\usepackage{iftex}

\ifpdf
  \usepackage{underscore}         % Only needed if you use pdflatex.
  \usepackage[T1]{fontenc}        % Recommended with pdflatex
\else
  \usepackage{breakurl}           % Not needed if you use pdflatex only.
\fi

\usepackage{float}

\usepackage{comment}
\usepackage{todonotes}

\title{EZASP - Facilitating the Usage of ASP}
\author{Rafael Martins
\institute{NOVA LINCS, NOVA University\\ Lisbon, Portugal}
\email{rlo.martins@campus.fct.unl.pt}
\and
Matthias Knorr
\institute{NOVA LINCS, NOVA University\\ Lisbon, Portugal}
\email{mkn@fct.unl.pt}
\and
Ricardo Gonçalves
\institute{NOVA LINCS, NOVA University\\ Lisbon, Portugal}
\email{rjrg@fct.unl.pt}
}
\def\titlerunning{EZASP - Facilitating the usage of ASP}
\def\authorrunning{Rafael Martins, Matthias Knorr \& Ricardo Gonçalves}
\def\copyrightholders{R. Martins, M. Knorr \& R. Gonçalves}
\begin{document}
\maketitle

\begin{abstract}
Answer Set Programming (ASP) is a declarative programming language used for modeling and solving complex combinatorial problems. 
It has been successfully applied to a number of different real-world problems.
However, learning its usage can prove challenging as the declarative language, from a conceptual perspective, differs substantially from imperative programming, and programs are not required to adhere to any particular structure, offering arguably almost too much freedom for a beginner.
Recently, a new methodology called Easy Answer Set Programming (Easy ASP) has been introduced that aims to aid in this learning process by focussing on a well-defined fragment of the ASP language and introducing additional structure to the programs.
However, while this methodology can indeed be employed, to the best of our knowledge, no tool integrates its features currently. 
In this paper, we present EZASP, a Visual Studio Code extension designed to support the development of ASP programs following the Easy ASP methodology.
It covers and extends the language fragment of Easy ASP and provides the user with warnings in the case of deviations from the methodology as well as the possibility to automatically reorder the program.
Complementarily, it also adds syntax error highlighting, including detection of non-safe variables directly while editing, and configurability, as all features can be optionally disabled.
A small user study in the context of university teaching suggests that these features are benefitial for both new and experienced users.
\end{abstract}

\section{Introduction}

Answer Set Programming (ASP) \cite{Lifschitz19} has emerged as a powerful approach for knowledge representation and reasoning, allowing users to model complex combinatorial (usually NP-hard) problems in a declarative manner.
ASP has been successfully applied in a number of real-world applications, such as scheduling, e.g., in the allocation of available personnel in the Gioia Tauro seaport \cite{ASPSeaport} and in the automated nurse scheduling at the University of Yamanashi Hospital \cite{ASPNurseScheduling}, planning and logistics, e.g., in the Partner Units Problem that involves organizing sensors and zones along the railway tracks to ensure safety \cite{ASPSiemens} and in the optimization of global aircraft assembly logistics at Airbus, where ASP is used to find valid supply chain configurations across multiple worldwide manufacturing sites \cite{ASPAirbus}, alongside many others (see, e.g., \cite{FalknerFSTT18}).

One of the frequent arguments in favor of the approach is its declarative language, which allows the succint specification of "what" the solution should be, instead of determining "how" the solution can be obtained as common in imperative programming languages.
In fact, unlike imperative languages, where a program is a sequence of instructions, an ASP program does not require any specific order for its composing pieces.
While this is certainly convenient for practitioners, using the language commonly proves challenging for newcomers due to the lack of structure in programs and the fundamental differences on how to create a program in the first place. 

Recently, the Easy Answer Set Programming methodology (Easy ASP) \cite{EasyASP} was proposed as a way to facilitate the learning and usage of ASP.
The main idea is to restrict to a fragment of the full ASP language that is amenable to creating programs with more structure and at the same time expressive enough to still capture common problems in ASP.
This is achieved on the one hand by defining a normal form of programs that aligns essential ASP constructs concisely with the well-established guess-define-test methodology, and, on the other hand, by organizing the program so that the relevant definitions and specifications are presented in a sequential and composing logical order.
It is argued that, following this methodology, beginners can more easily learn how to create ASP programs.

While employing this methodology can be done manually, for beginners, this can be time-consuming and error-prone.
Thus, it would be convenient to have automated support for it. 
Yet, no tool exists for this purpose.
In fact, IDEs for ASP are not widespread, as, in the limit, a simple text editor suffices, even if if lacks a number of useful features a developer could benefit from.
Notable exceptions to the lack of IDEs are SeaLion \cite{OetschPT18}, a standalone desktop tool that includes advanced debugging features, but it seems to have limited support nowadays; LoIDE \cite{GermanoCP18}, a web-based IDE with ongoing development,\footnote{\url{https://demacs-unical.github.io/LoIDE/}} ASPChef \cite{AlvianoRF25}, a web application, aiming to facilitate visualizations and processing results of ASP programs,\footnote{\url{https://asp-chef.alviano.net/}} and, more basic in comparison, the general-purpose IDE Visual Studio Code\footnote{\url{https://code.visualstudio.com/}} with a couple of extensions for syntax-highlighting and facilitating the execution of ASP programs directly from the editor.
Still, none of these options focus on aiding beginners specifically.

In this paper, we introduce EZASP,\footnote{\url{https://github.com/RafaelMartins4/EZASP-2}} an easy-to-install extension for Visual Studio Code, a tool widely used for a number of different programming languages.
Our extension allows the application of the EasyASP methodology, extended here to capture a wider class of programs, within an IDE for the first time, including a number of features that aid both new and experienced users, ultimately facilitating the adoption of ASP.
Its contributions can be summarized as follows:
\begin{itemize}
	\item Building on a parser developed for the extension, syntax errors are detected and appropriately highligthted within the edition pane, facilitating the early detection of problems;
	\item Taking into account an adjusted ordering of the different constructs in ASP, errors on this order, as well as on the composition of the program in terms of their logical structure, are identified and highlighted, alerting to misalignments with the EasyASP methodology;
	\item If the user wishes, the current program can be automatically re-ordered so that it aligns with the intended structure according to EasyASP, keeping the structure of the program otherwise intact as much as possible;
	\item All these features can be activated/de-activated in a configuration file (in json), making it possible to easily tailor the user experience to their expertise or preferences. 
\end{itemize}

The remainder of the paper is structured as follows.
We first recall necessary notions on ASP and the methodology of Easy ASP in Sections~\ref{sec:asp} and \ref{sec:easyasp}, respectively.
Then, in Section~\ref{sec:ezasp}, we discuss in detail our new extension and all the included features.
We also present a preliminary study with students on the usability and benefits of this tool in Section~\ref{sec:tests}, before we conclude in Section~\ref{sec:concl}. 

\section{Answer Set Programming}\label{sec:asp}

We briefly recall relevant notions and notation for logic programs under the answer set semantics.
Here, we follow standard definitions, but use a syntax format closely aligned with the program aspect.

A logic program $P$ is a finite set of rules $r$ of the form:
\begin{equation}\label{rule}
	a_0 :- a_1,...,a_m,\: not\: a_{m+1},...,\: not\: a_{m+n}.
\end{equation}

\noindent where each \(a_i\) ( \( 0 \leq i \leq m+n \)) is an \emph{atom}. An atom takes the form \(p(t_1,...,t_k)\), where \emph{p} is a predicate symbol of arity \emph{k}, and \(t_1,...,t_k\) are terms, built from variables and constants of the (implicit) language of $P$. We define as literals both an atom \(a\) as well as its negation,  \(not\: a\). Atoms to the left of the 'if' symbol (\(:-\)) of a rule \emph{r} are said to be the head of the rule,\footnote{In general, disjunctions are permitted in the head, but we simplify here for the sake of presentation.} and we write \(head(r) = a_0 \), whereas the literals to the right of the arrow are the body of the rule (also known as body literals), and we write \(body(r) = \{ a_1,...,a_m,\: not\: a_{m+1},...,\: not\: a_{m+n} \}\).
We may omit $r$ from $body(r)$ if clear from the context.

We call rules with $n=0$ \emph{definite}, rules with $m=n=0$ \emph{facts}, and admit that $a_0$ is $\bot$ to represent \emph{constraints}, commonly omitting $:-$ in the second case and $\bot$ in the final one. 

The ground instantiation of a program, $ground(P)$ is obtained by replacing all (first-order) variables with constants from the given language of the program in all possible ways.
To ensure that grounding results in a finite set of rules \emph{variable safety} is employed.
Technically, it requires for a rule of the form \ref{rule} that each variable in it occurs in at least one non-negated literal in the body.
A program $P$ is \emph{safe} if all its rules are safe.
%Variable safety is a mechanism that verifies that all variables in a program can be grounded by the system. This mechanism ensures that each variable has a finite set of possible values, thereby guaranteeing that the program is fully groundable and, consequently, executable. If an ASP program contains a variable with an infinite domain, it becomes unsolvable and, as a result, produces no answer sets.
A set of ground atoms $S$ is then an answer set of $ground(P)$ if $S$ is the subset-minimal model of the reduct $ground(P)^S$ obtained as $\{	a_0 :- a_1,...,a_m\mid r$ of the form (\ref{rule}) in $ground(P)$ and $\{a_{m+1},\ldots,a_{m+n}\}\cap S = \emptyset\}$.

ASP also admits a number of additional language constructs which will be important in the remainder of the paper, and we recall these in the following.

\paragraph{Constant Declaration}

The \texttt{const} directive allows for the definition of values that, once set, are fixed throughout the program. %They act as a symbolic placeholder for specific values that cannot be changed when running the program. 
In clingo \cite{clingo}, a constant declaration is written as: \(\#const \; \; constant\_name \; = \; value.\)
Here, \(constant\_name\) is a string starting with a lowercase letter, and \(value\) is the fixed value that is assigned to the constant. 
The value can be re-set on program call.
%After assigning a value to a constant, the user can use the constant name as a replacement to the fixed value, leading to more readable and more easily modifiable code.

\paragraph{Choice Rules}

In ASP, choice rules are used to non-deterministically select subsets of atoms to be true in a solution. This is especially useful to model problems that involve optional selections
or combinations of a subset of elements. The structure of choice rules is as follows:  

\[lb\{a_1,\,a_2, ..., a_n : b_1,\,b_2, ..., b_k \}ub :- \; body.\]

%Here:
\begin{itemize}
  \itemsep0em
  \item[-] \(a_1,\,a_2, ..., a_n\) represents the head of the choice, which denotes the atoms that may be selected.
  \item[-] \(b_1,\,b_2, ..., b_k\) represents the condition of the choice, which denotes an expression that must be true in order to select atoms from the head; more generally, together, the head and the condition of the choice may also be an expression that can be instantiated into a set of ground atoms.
  \item[-] \(lb\) (optional) represents the lower bound that specifies the minimum number of atoms to be included in any solution.
  \item[-] \(ub\) (optional) represents the upper bound that specifies the maximum number of atoms to be included in any solution.
  %\item[-] \(body\) is the usual rule body, representing the condition under which the choice is applied.
\end{itemize}
A choice rule with an empty body is called \emph{choice}.
Often, also the term \emph{cardinality constraint} is used to refer to choice rules with bounds, while the term choice rules is reserved for those without.\footnote{Choices \(lb\{a_1,\,a_2, ..., a_n : b_1,\,b_2, ..., b_k \}ub\) may also occur in the body of a rule as a representation of part of the condition.}

\paragraph{Optimization Statements}

Optimization statements enable the user to specify preferences among the valid solutions based on a given set of criteria. These statements allow the user to define what is the prefered 
solution by providing a way to rank solutions through the use of costs and priorities tied to specific conditions or configurations. The structure of an optimization statement is 
the following:
  \[\#directive \, \{w_1@p_1, t_1 : L_1; w_2@p_2, t_2 : L_2; ...\}.\]
\begin{itemize}
  \itemsep0em
  \item[-] \textit{directive} is the optimization objective (minimize, maximize).
  \item[-] \(w_i\) is the weight associated with the condition \(L_i\), a literal or a complex condition.
  \item[-] \(p_i\) is the priority level of the condition in comparison to other conditions. Higher priority will be optimized first, and lower priorities only are considered in case of a tie. 
  \item[-] \(t_i\) are optional terms that serve as labels for specific instances of the literals \(L_i\).
  \item[-] \(L_i\) is a literal that triggers the associated cost when true (again, can be a complex expression, instantiated into different specific cases).
\end{itemize}

It is well-known that optimization statements can also be represented using weak constraints of the form \(:\sim body. \; [w \; @ \; p, t]\). %, where \(body\) can be a single literal \(L\).

\paragraph{Show Statement}

Show Statements are used to control the output of the program by specifying which atoms should be printed in the result, instead of displaying simply all true atoms. 
For that, the user must specify the atoms to be printed indicating for each predicate a line:
  \[\#show \; \; predicate\;/\;arity.\]
%Here, 
%\(arity\) represents the number of arguments that predicate takes.

%\paragraph{Variable Safety}

\section{Easy Answer Set Programming}\label{sec:easyasp}

The Easy Answer Set Programming (Easy ASP) methodology was proposed by \textit{Fandinno et al.}~\cite{EasyASP} as a structured approach to programming in ASP. Its main objective is 
to make ASP more accessible for beginners by enforcing structural guidelines that promote a standardized organization of programs.
This builds on two essential components: one is a fragment of ASP, called austere ASP, which to some extent limits the expressiveness, but maintains the capability to represent problems, which, in turn, can be used to define a structured approach to write programs. 
The second refers to the idea that, usually, ASP programs are somewhat compositional, in that certain concepts are defined in a sequential/hierarchical way, and the idea is to ensure that this logical structure is preserved in the organization of the program itself, aligned with the ideas of stratification \cite{AptBW88}.
We next briefly describe these two ideas in more detail.

\paragraph{Austere ASP}

Austere ASP \cite{EasyASP} is a normal form for ASP programs, decomposing them into four sets of constructs: Facts, Choices, Definite Rules and Integrity Constraints. This organization separates the different roles that constructs play in the program and aligns with the guess-define-test paradigm that is adopted by Easy ASP. As a result, Austere ASP is integrated into the Easy ASP methodology as a way to facilitate the development of ASP programs by enforcing the ordering of the program:

\begin{itemize}
  \itemsep0em
  \item \textbf{Facts:} These define the known information and the problem instance. They are unconditional statements about the domain of the problem, and may change to represent another instance.
  \item \textbf{Choices:} These rules (without body) are used to guess models non-deterministically. Note that in Austere ASP, these rules are not allowed to mention cardinalities to separate these two concerns, as such restrictions can still be expressed in integrity constraints without losing expressiveness.
  \item \textbf{Definite Rules:} These are deterministic rules with the objective of introducing auxiliary concepts based on the guessed solutions, allowing us to derive additional properties or relationships.
  \item \textbf{Integrity Constraints:} These define the problem's constraints, i.e., undesired properties for solutions, that are used to eliminate invalid solutions.
\end{itemize}

This structure aligns naturally with the guess-define-test paradigm: facts define the problem instance, choices generate candidate solutions, definite rules derive relevant properties and relationships, and integrity constraints ensure that only valid solutions are retained.
The definite rules also ensure that, once solutions are guessed in choices, models can be computed deterministically as definite rules (without negation) do not introduce any ambiguities \cite{EasyASP}.

To be able to nevertheless use negation in the defining part without losing this determinism, Easy ASP also imposes stratification as a structural guideline. 
Before we briefly discuss this, we recall the notions of dependencies between predicates and definitions of a predicate in a program.
Namely, a predicate \(p\) depends on a predicate \(q\) if \(q\) is present in the
body of a rule that defines \(p\), i.e., $p$ occurs in its head.
The \emph{definition} of a predicate \(p\) is the subset of program \(P\) consisting of all rules with \(p\) in their head.

\paragraph{Stratification}

Stratification refers to organizing the program's predicates into layers such that any given predicate only depends on predicates from the current or previous layers. This formatting guarantees a logical progression when deriving answer sets as well as a more structured organization of the program itself. The objective of stratification is to prevent problematic scenarios that might appear through predicate recursion. 
In some cases, the lack of answer sets may also be an indicator for errors in the specification of the problem, which can then lead to inconsistent results.
To capture this formally, the following notions on predicate dependencies are defined:
\begin{itemize}
  \itemsep0em
  \item If a predicate \(p\) depends negatively on predicate \(q\) (i.e. \(p :- \; not \; q\)), then \(q\) must be defined in a lower layer than \(p\).
  \item If \(p\) depends positively on predicate \(q\) (i.e. \(p :- \; q\)), then \(q\) can be defined in the same layer as \(p\) or in a lower layer.
\end{itemize}
A program is then stratified, if its rules can be partitioned into layers such that the definitions of predicates abide with these two notions \cite{AptBW88}.

Not all programs are stratified, e.g., \(p :- \; not \; q\) and \(q :- \; not \; p\), which introduces a non-deterministic choice between \(p\) and \(q\).
Other cases of non-stratified programs may have no answer sets, i.e., they behave like a constraint in a certain way, which thus again mixes different concerns in terms of modelling.
In each case, this is caused by negative dependencies, and to avoid these problems altogether it is assumed that these rules together be stratified, i.e., they do not admit such cyclic dependencies involving negation.
This indeed allows to use stratified rules with negation instead of definite ones in austere ASP, while preserving the determinism of answer set computation once choices have been fixed.

Altogether, \emph{Easy logic programs} are defined as a sequence of facts, choices, a stratified partition of rules with negation, called \emph{definitions}, and constraints.
This provides the desired structure of programs and allows defining the Easy ASP methodology by compartmentalizing a program into these parts \cite{EasyASP}.

\section{EZASP}\label{sec:ezasp}

In this section, we present EZASP, a Visual Studio Code (VS Code) extension\footnote{https://github.com/RafaelMartins4/EZASP-2} developed with the objective of implementing the Easy ASP methodology in a practical manner. It provides a structured approach
to ASP by enforcing the principles of the Easy ASP methodology through the use of errors and warnings built into the code editor. This allows users to easily identify and correct 
instances where the ordering and stratification principles of the Easy ASP methodology are not being followed, ultimately leading to more organized and intuitive ASP programs. Additionally, EZASP offers
several other features that further enhance the user experience and facilitate the development of ASP programs. 

\subsection{EZASP Dependencies and Configuration File}

EZASP is built using JavaScript and includes two other VS Code extensions: the \textit{Answer Set Programming Language Support} extension,\footnote{https://github.com/CaptainUnbrauchbar/ASP-Language-Support} which is responsible for integrating the clingo solver into the Visual Studio Code environment that is then used to compute the answer sets of ASP programs, and the \textit{Answer Set Programming syntax highlighter} extension,\footnote{https://github.com/ArnaudBelcour/asp-syntax-highlight} which provides complementary syntax highlighting for ASP programs inside Visual Studio Code. Both of these extensions are automatically installed when installing EZASP, ensuring all users have access to the  available convenient tools within VS Code when installing EZASP.

Furthermore, the \textit{Answer Set Programming Language Support} extension provides a configuration file that allows users to customize the behavior of the extension, such as startup arguments, configuring multiple file programs, defining solving limits and more. EZASP also leverages this configuration file to allow users to enable or disable any of its features, giving them the flexibility to customize their development experience according to their preferences and needs. To create and utilize this file, EZASP provides a VS Code command that generates a default configuration file in the current directory, which users can then edit to affect any ASP programs that share the same directory.

\subsection{Syntax Checking}
We start with the EZASP feature of syntax checking, which is an essential feature, useful for any ASP developer, regardless of their experience level. 
EZASP implements syntax checking for ASP programs through the use of a custom-built parser that analyzes the code and identifies syntax errors. To do this, EZASP utilizes the
ANTLR (ANother Tool for Language Recognition) \cite{ANTLR} parser generator, which allows the specification of the accepted language through the use of a grammar file. 
Using a carefully defined grammar, EZASP is able to closely align with the syntax of clingo, therefore being able to provide real-time feedback to users, by highlighting syntax errors directly in the code editor.

Additionally, the ANTLR parser allows EZASP to be able to extract relevant information from each construct in the program, such as which predicates are used in each rule and their 
exact position in the program. As we will see later, this information is crucial for implementing other EZASP features. %, as it allows for a detailed analysis of the program's structure.

\paragraph{Syntax Errors}

To capture syntax errors present in ASP programs, EZASP utilizes a custom error listener that stores the syntax errors recognized by ANTLR. Along with each error, ANTLR generates additional relevant information, which is also stored by the listener, such as the offending symbol, i.e., the character that prompted the parser to mark it as an error, and the corresponding line and column positions. This information, together with an automatically generated error message produced by the ANTLR parser, is then used to create the syntax error instances that are highlighted in the code editor. 

\begin{figure}[t!]
    \centering
    \includegraphics[scale=0.85]{Figures/ezasp_syntaxError.png}
    \caption{Syntax Error examples in EZASP}
    \label{syntaxError}
\end{figure}

As shown in Figure \ref{syntaxError}, to visually highlight syntax errors in the code editor, EZASP underlines a predefined range of five characters centered at the offending symbol. This approach ensures that errors remain clearly visible even in larger programs while maintaining a clean editor interface and avoiding excessive visual clutter. Additionally, two notable edge cases require special handling: when a syntax error occurs at the start or at the end of a line. 

For errors at the start of a line, EZASP examines the previous line (if it exists) to determine whether part of the underline should extend to that line. Specifically, EZASP checks how the previous line ends:

\begin{enumerate}
  \item If the previous line ends with a statement terminator, it indicates a correctly closed statement. In this case, the underline remains within the current line and is extended forward to maintain the predefined underline length.
  \item If the previous line does not end with a terminator, the statement may be incomplete. Therefore, part of the underline is applied to the end of the previous line while ensuring that the total underline length remains consistent.
\end{enumerate}

For errors at the end of a line, a similar approach is applied. EZASP starts by verifying if the current construct ends with a statement terminator. If yes, the underline is extended backward so that its total length matches the predefined range. Otherwise, we cascade the error to the next line.

\subsection{Unsafe Variable Detection}

Another important feature present in EZASP is the detection and highlighting of unsafe variables in ASP programs. Unsafe variables are a common error in ASP programming. 
This error occurs when a variable cannot be grounded by the system, meaning that it cannot be assigned a finite set of values. This results in the program being unsolvable by ASP
systems and, therefore, it is crucial for developers to identify and correct these issues in their program.

This kind of error is not common in other programming languages, and it can therefore be difficult for newcomers to understand it and how to identify it. To aid with that, EZASP implements a real-time detection system capable of capturing the same unsafe variable errors as the clingo solver.

For this, EZASP utilizes three structures that are populated during the parsing process for each rule: 

\begin{itemize}
  \item \textbf{totalVariables}: holds all the variables present in the rule that is being parsed.
  \item \textbf{groundedVariables}: contains all the variables that are grounded by the rule that is being parsed.
  \item \textbf{linkedVariables}: stores pairs of variable sets that are linked together. It is used to ground new variables if all variables in one of the sets are grounded.
\end{itemize}

As variables in different rules are always different, variables must always be grounded within the scope of their own rule. The exception to this are variables that occur in constructs that introduce a more restricted local context inside the rule itself, such as choices and aggregate atoms. In these cases, a variable may be grounded either within the smaller context of 
the atom itself or by another construct in the rule's global context, which grounds the variable for the rule as a whole.
In these cases, EZASP utilizes similar structures to capture the same information for variables within a certain context in a rule. %This is useful to detect unsafe variables inside constructs such as choices, where variables can be grounded on only a part of the rule. 

As a result, EZASP creates \textit{contextVariables} set, \textit{contextGroundedVariables} set and \textit{contextLinkedVariables} array whenever necessary, and populates these by adding a new element whenever a new variable is encountered during the parsing of a rule.

To populate the \textit{groundedVariables} and \textit{contextGroundedVariables} sets, EZASP must check a variety of cases that can ground variables in ASP programs. 
To give an idea of this variety necessary to be covered, we provide a non-exhaustive list of occurrences that allow grounding variables in ASP programs and which goes beyond the well-known standard definition in Section~\ref{sec:asp}:

\begin{itemize}
    \item In a positive atom in the rule's body;
    \item In a comparison. A variable is grounded in this case only if (in accordance with the behavior of clingo): the comparison occurs in the body of the rule and the comparator denotes equality (i.e. '=', '==' or 'not ... !=') or the comparison occurs in the head of the rule and the comparator denotes inequality (i.e. '!=' or 'not ... =='). Furthermore, the comparison must include variables on only one side (e.g. 'p(X) == 1' grounds X, whereas 'p(X) == Y' does not ground either variable); 
    \item In the condition of a choice. This grounds the variable within the context of the choice itself. In addition, this also grounds over occurrences of the same variable in the rule.
\end{itemize}

Note that \textit{linkedVariables}, and its construct-specific counterpart, \textit{contextLinkedVariables}, store pairs of variable sets that are linked, meaning that if 
all variables in one of the sets are grounded, then all of the variables in the corresponding set must be grounded as well. These arrays are populated whenever the rule contains 
a comparison with variables on both sides, as long as the comparator denotes equality (for comparisons in the body of a rule) or inequality (for comparisons in the head of a rule).
This mechanism operates in this way because unsafe variables are variables with an infinite set of possible values. When we have two sets of variables that are linked together through a comparison, if one of the sets contains only variables with finite domains (i.e. are grounded), then we can infer a finite set of values for the variables on the other side of the comparison, which allows the solver to ground them as well.

After parsing each rule, EZASP verifies unsafe variables by checking if there are any variables in the \textit{totalVariables} set that are not present in the \textit{groundedVariables} set. If this is the case, EZASP creates an error instance that lists all the unsafe variables found in each rule and highlights them in the code editor. For variables in the \textit{contextVariables} set, a similar process is applied, but EZASP must also check whether the variables are grounded within the context of the construct where they are used. As a result, besides checking the \textit{groundedVariables} set, EZASP must also check the corresponding \textit{contextGroundedVariables} set.

\subsection{Ordering of the Program}\label{subsec:order}

As mentioned previously, EZASP implements the ordering proposed by the Easy ASP methodology, which enforces a specific order of constructs within ASP programs to promote a more structured way of writing ASP programs.

The Easy ASP methodology proposed the following order of constructs: Facts, Choices, Definitions (as stratified rules), and Integrity Constraints. 
However, while this covers a number of additional constructs in the ASP language that can be rewritten into a combination of the former \cite{EasyASP}, other constructs are not covered.
We therefore have extended the permitted kind of constructs in a program, admitting also those discussed in Sec.~\ref{sec:asp} that are not part of the Easy ASP program structure.
In addition, we also generalized choices to choice rules including cardinality constraints.
While one could argue that this deviates to some extent from the envisioned approach, from our experience in teaching in undergraduate and graduate courses using ASP, these additional constructs are convenient to have early on, so we decided to incorporate them.\footnote{Indeed, all but the change from choices to choice rules are complementary additions to the core of the order of Easy ASP.}
The final order of constructs considered by EZASP is as follows:

\begin{itemize}
  \itemsep0em
  \item Constant Declarations.
  \item Facts.
  \item Choice rules.
  \item Definitions.
  \item Constraints.
  \item Optimization Statements.
  \item Show Statements.
\end{itemize}

In order to enforce this order, EZASP parses each construct in the program and assigns it one of the aforementioned categories. Then, with a list of all parsed constructs, EZASP checks if they are in the correct predefined order. If any construct is found to be out of order, EZASP creates a warning instance that is displayed in the editor through a yellow underline, and a warning message indicates the expected category for the construct that is not aligned.

\subsection{Stratification}

EZASP also implements Stratification as proposed by the Easy ASP methodology, which divides ASP programs into layers based on predicate dependencies to aid with creating a logical progression within programs and avoid certain problems that can arise from predicate recursion involving negation.

To enforce this, EZASP keeps track of the locations where each predicate is defined and where it is used in the program. This information is crucial to determine the dependencies between predicates and to identify any violations of the stratification rule. As a result, EZASP's parser is designed to extract this information during the parsing process and store it in two map structures that associate each predicate (and arity) with its definition and usage locations in the program. With this information, EZASP verifies the stratification of the program by guaranteeing that all predicates are defined in a layer that is lower than the layers where it is used, i.e., it is defined in the program before it is used.
 
If any violation of the stratification rule is detected, EZASP creates a warning instance if the predicate is used before it is defined or an error instance if the predicate is used but never defined in the program. Both instances are displayed in the code editor through yellow and red underlines, respectively, and are accompanied by a warning or error message that indicates the nature of the problem.
This also takes advantage of the configuration file if a solution is created using several files: it checks the other files under the assumption that these have been defined before to ensure that within the individual file no misleading error messages are displayed.

As an additional benefit, this also allows the detection of typos and missing arguments in an atom of a predicate, as any of those are displayed automatically as not defined, which we believe is convenient for any user to have. Also, note that ASP programs that have a group of predicates that positively depend on each other (e.g., \(p :- q\) and \(q :- p\)) will always present stratification warnings, as no ordering of these predicates can satisfy the stratification rule.

\subsection{Automatic Code Reordering}

Writing programs that follow the Easy ASP methodology can be demanding, especially for beginners who may not be fully familiar with the correct ordering of all constructs yet and when faced with programs of increasing size.
To aid the user beyond guidance by the presented warnings, EZASP offers an automatic code reordering mechanism that aims to restructure the program so that it aligns with the desired order of rules and stratification.
This convenient functionality is available through a button in the editor that is visible whenever an ordering warning is present in the program.

\paragraph{Reordering Logic and Strategy}

To implement this feature, EZASP employs an algorithm that is designed to reorder constructs in a way that respects the order described in Section~\ref{subsec:order} as well as stratification, while minimizing the number of changes made to the original program. The algorithm is divided into four phases: dividing the program into reorderable blocks, ordering constructs, minimizing stratification warnings and applying results. 

Firstly, EZASP divides the program into blocks that are then reordered as a whole. This is necessary to ensure that comments are not displaced or overwritten by other constructs during the reordering process. To achieve this, EZASP extends each construct's range both upward and downward. Upward extension captures comments between the current construct and the previous one. Downward extension includes line comments that follow the construct and block comments that begin after it (regardless of the block comment size), provided that no other construct appears between the construct and these comments. We are left with a list of blocks that can be reordered without the risk of losing information. This also ensures that comments maintain their relative position to the constructs they refer to in the resulting program, which is crucial for readability and understanding.

After partitioning the program this way, EZASP reorders constructs to match the intended ordering proposed. For this purpose, EZASP creates a constructsByType object that stores an array for each of the possible construct types. The array is sorted based on the 
order in which each construct appears in the original program. As a result, any construct that is out of order in the original program is moved to the extremities of
the array corresponding to their type. This method of reordering ensures that we minimize the number of changes made to the original program,
moving each out-of-order construct to their corresponding group, without changing anything that already appears in the right order.

Then, EZASP tries to minimize the number of stratification warnings present in the final program. To do this, EZASP implements a sorting strategy that is applied to the critical sections of the program, i.e., those that contain constructs that can both define and use predicates (facts, choice rules and definitions). All other sections are limited to either defining or using predicates, so no possible reordering within these sections would resolve stratification warnings without moving the construct out of its designated section, which is already fixed in the previous step. 

The sorting strategy implemented by EZASP includes three steps: determining the predicates that each construct defines and uses, build a dependency graph, and execute a topological sort of the graph.
This uses dependencies in the sense defined before but on the level of constructs, i.e.,  
construct \texttt{A} depends on construct \texttt{B} if construct \texttt{A} uses a predicate that is defined by construct \texttt{B}. As a result, construct \texttt{A} must appear after construct \texttt{B} in the final sorting result. 

To understand the dependencies between constructs, EZASP utilizes the information extracted during the parsing process. For each construct, we have information about the predicates it defines and the predicates it uses. With this information, we can determine the dependencies between constructs and build a graph that represents these dependencies. In this graph, each node represents a construct and each directed edge represents a dependency from the construct that uses a predicate to the one that defines it. Finally, to achieve a final sorting that minimizes stratification warnings within a section, EZASP executes a depth-first search topological sort of the dependency graph. This sorting algorithm guarantees that each construct is placed after the constructs that define the predicates it uses. Furthermore, if there are any cycles present in the graph, it indicates that there are stratification warnings that cannot be resolved through reordering, and a special warning message is triggered to inform the user about this issue.

After sorting the critical sections, EZASP applies the results of the reordering to the code editor by iterating the resulting array and utilizing the constructs ranges to gather the corresponding text in the original program. The implementation also guarantees that the resulting program contains a newline token between each code block, as to create a less cluttered program.

\section{Usage Tests}\label{sec:tests}

To evaluate the effectiveness of EZASP in facilitating the development of ASP programs, we conducted a small user study. Participants were students, some from undergraduate level and a larger share from graduate level course, each course exposing students to ASP for the first time. Though the graduate level course includes a considerable share of students who had some experience with ASP from the brief contact in the undergraduate course one or two years before. Participation was entirely voluntary and anonymous.
A questionnaire was provided where participants were asked to quantitatively evaluate every feature of the extension. The questionnaire was answered by 24 participants, where 18 had prior experience with ASP and 6 did not. Participants were asked to evaluate the usefulness of each feature on a scale from 1 to 5, where 1 represents 'Not Useful at all' and 5 represents 'Very Useful'. Additionally, some features were also evaluated based on their ease of usability, with a similar scale as the one used for usefulness. Furthermore, participants were also given the opportunity to provide qualitative feedback and suggestions for each feature, which can be used to motivate future improvements to the extension.

\subsection{Syntax Errors}

The syntax error highlighting feature was evaluated through two components: error messages and visual underlining within the editor. Error messages were rated with an average score of 3.79, with 62.5\% of participants considering them 'Useful' or 'Very Useful'. In contrast, the visual underlining performed better, reaching an average score of 4.21 and a positive rating from 79.2\% of respondents.

This indicates that, while users benefit from the visual presentation of syntax errors in the editor, the error messages themselves are sometimes less effective in identifying the precise cause of errors.

\subsection{Unsafe Variable Detection}

Participants rated the unsafe variable detection mechanism with an average usefulness score of 4.08, and 75\% classified it as 'Useful' or 'Very Useful'. These results suggest that the unsafe variable detection mechanism is generally considered useful by users, which is expected given that unsafe variables are a common error in ASP programming and can be difficult to identify without expertise or proper tools.

\subsection{Stratification Issues}

The stratification mechanism received particularly strong feedback across both of its evaluated aspects. The visual highlighting of stratification issues achieved one of the highest ratings overall, with an average of 4.46 and 87.5\% positive responses. The distinction between warnings and errors was also well received, with an average rating of 4.08 and 66.7\% of participants finding it useful.
These results highlight the importance of clearly communicating both the presence and severity of stratification problems. While both elements contribute positively, the especially high rating of visual highlighting suggests that immediate, in-editor feedback is particularly valuable to users, also to help detect errors in predicate names or their arity.

\subsection{Automatic Code Reordering}

To evaluate the automatic code reordering button, participants were asked to rate the usefulness of the feature and its ease of use, based on how easy it is to find and utilize this feature. Participants rated the usefulness of the automatic code reordering feature with an average of 4.46, with 7 participants rating it as 'Useful' and 15 participants rating it as 'Very Useful'. On the other hand, the ease of use of the feature received an average rating of 3.37, with only 45.8\% of participants rating it as 'Useful' or 'Very Useful'. These results suggest that while the automatic code reordering feature is very appreciated by users, it was often difficult to find, as commented in the detailed feedback, which indicates that the automatic code reordering button may require a more visible or intuitive placement in the code editor.

\section{Conclusion}\label{sec:concl}

We have presented EZASP, an extension for Visual Studio Code that provides a comprehensive set of functionalities designed to support the practical application of the Easy ASP methodology within a widely adopted IDE. Beyond Easy ASP-specific features, EZASP introduces quality-of-life (QoL) improvements that simplify ASP development as a whole and contribute to the production of correct ASP programs for both beginners and experienced users. The implemented functionalities were evaluated overall positively by users with varying levels of ASP experience, providing also indications for possible improvements. %allowing quantitative and qualitative feedback from different perspectives. As shown in Section \ref{sec:tests}, the results of the evaluation were largely positive, with the majority of participants classifying the functionalities as useful. Furthermore, the qualitative feedback from participants allowed for the identification of future improvements to the extension, such as revisiting the placement of the code reordering button.

Still, EZASP presents some limitations when compared e.g., to the parser of state-of-art solver clingo. One such limitation is the absence of type-incompatibility detection that happens, for example, when a user defines an arithmetic operation between two constructs that are not compatible: for example, '\texttt{q(X+Y) :- X == 1, Y == "str".}'. In this case, clingo throws an `operation undefined' error, whereas EZASP currently fails to identify the problem.

Future work should therefore focus on aligning EZASP's behaviour more closely with that of clingo, as well as extending its set of functionalities to further support and simplify the usage of ASP for users across different levels of ASP experience.

\section*{Acknowledgments}

We thank the anonymous reviewers for their valuable feedback.
We acknowledge support by project BIO-REVISE (2023.13327.PEX) %, project FAIR (2024.16987.PEX), 
and NOVA LINCS (UIDB/04516/2025) with the financial support of FCT.IP.

\bibliographystyle{eptcs}
\bibliography{generic}
\end{document}